\documentclass[aps,prc,twocolumn,showpacs,showkeys,preprintnumbers,amsmath,amssymb,a4paper]{revtex4-1}

\usepackage{amssymb}
\usepackage{amsmath, bbm}
\usepackage[figuresright]{rotating}
\usepackage{capt-of}

\newcommand{\nb}{\bar n}
\newcommand{\osc}{n-\bar n}
\newcommand{\npbar}{\bar np}
\newcommand{\nnbar}{\bar NN}
\newcommand{\ppbar}{\bar pp}

\newcommand{\be}{\begin{eqnarray}}
\newcommand{\ee}{\end{eqnarray}}

\newlength{\feynwidth} 
\newlength{\feynwidthbig} 

\usepackage{xcolor} 

\begin{document}

\title{Neutron-antineutron oscillations in the deuteron studied with $NN$ and $\nnbar$
interactions based on chiral effective field theory}

\author{Johann Haidenbauer$^{1}$, Ulf-G. Mei{\ss}ner$^{2,1,3}$}

\affiliation{
$^{1}${Institute for Advanced Simulation, Institut f\"ur Kernphysik and J\"ulich Center for Hadron Physics,
  Forschungszentrum J{\"u}lich, D-52425 J{\"u}lich, Germany}
\\
$^{2}${Helmholtz-Institut f\"ur Strahlen- und Kernphysik and Bethe Center for Theoretical Physics,
  Universit\"at Bonn, 53115 Bonn, Germany}
\\
$^{3}${Tbilisi State  University,  0186 Tbilisi, Georgia}
}
\begin{abstract}
Neutron-antineutron ($n-\bar n$) oscillations in the deuteron are considered. Specifically, 
the deuteron lifetime is calculated in terms of the free-space $n-\bar n$ oscillation time 
$\tau_{n-\bar n}$ based on $NN$ and $\bar NN$ interactions derived within chiral effective 
field theory (EFT). This results in $(2.6\pm 0.1) \times 10^{22}\,\tau^2_{n-\bar n}$~s, 
which is close to the value obtained by Dover and collaborators more than three decades 
ago, but disagrees with recent EFT calculations that were performed within the perturbative 
scheme proposed by Kaplan, Savage, and Wise. 
Possible reasons for the difference are discussed. 
\end{abstract}


\maketitle

\section{Introduction}

Neutron-antineutron ($\osc$) oscillations involve a change of the baryon number ($B$)
by two units ($|\Delta B| = 2$). An experimental observation would allow a glimpse on 
physics beyond the standard model, see e.g.~\cite{Mohapatra:2009}.  
Since in such oscillations $B$ is violated the process satisfies one of the
Sakharov conditions~\cite{Sakharov:1967} that have been formulated in 
order to explain the observation that there is more matter than anti-matter in 
the universe.
Given such important and far-reaching consequences it is not surprising that there 
is a vast amount of literature on this topic \cite{Phillips:2014}. 
Indeed, even within the past two years a wealth of papers have been published
that deal with various and quite different aspects of $\osc$ oscillations  
\cite{Bijnens:2017,Gardner:2018,Grojean:2018,Oosterhof:2019,Rinaldi:2019,Gardner:2019,Berezhiani:2019,Kerbikov:2019}. 

The key quanitity in this subject is the free $\osc$ oscillation time, $\tau_{\osc}$. 
The presently best experimental limit on it is $\tau_{\osc} > 0.86 \times 10^8 \ {\rm s} 
\approx 2.7$ yr (with 90\,\% C.L.) \cite{BaldoCeolin:1994}. 
Additional information can be gained by studying $\osc$ oscillations in a nuclear 
environment. Corresponding experiments have been performed, e.g., for 
$^{56}$Fe \cite{Chung:2002}, $^{16}$O \cite{Abe:2015}, and for the deuteron ($^{2}$H) 
\cite{Aharmim:2017}, while others are planned \cite{Barrow:2019}. In such a case the 
oscillation process is suppressed as compared to the free situation. The pertinent 
lifetime $\tau_{\rm nuc}$ is commonly expressed in terms of the one in free space as 
\cite{Phillips:2014}
\begin{equation}
\tau_{\rm nuc}=R\, \tau^2_{\osc} \ , 
\end{equation}
where $R$ is an intranuclear suppression factor, also called reduced lifetime, that 
depends on the specific nucleus. It can be calculated from nuclear theory and then 
can be used to relate the measured lifetimes of 
those nuclei with the free $\osc$ oscillation time \cite{Phillips:2014}, see, e.g., 
Refs.~\cite{Sandars:1980,Dover:1983,Alberico:1991,Hufner:1998,Friedman:2008}. 

For a long time the suppression factors published in Ref.~\cite{Dover:1983} have been 
used as standard by experimentalists in the interpretation of their measurements
\cite{Chung:2002,Aharmim:2017}. For example, in case of the deuteron the corresponding
value is $R \sim (2.40-2.56) \times 10^{22}$ s$^{-1}$, a prediction based on the 
phenomenological antinucleon-nucleon ($\nnbar$) potentials by Dover and
Richard~\cite{Dover:1980,Dover:1982}. 
Recently, however, those values have been called into question in a work by Oosterhof
et al.~\cite{Oosterhof:2019}. In that study an effective field theory for the 
$|\Delta B| = 2$ interaction is constructed and the quantity $R$ is evaluated within the 
power counting scheme proposed by Kaplan, Savage, and Wise (KSW) \cite{Kaplan:1998,Kaplan:1999}   
for the nucleon-nucleon ($NN$) and $\nnbar$ interactions. The value of $R$ for the deuteron 
obtained in that approach is with $(1.1\pm 0.3) \times 10^{22}$ s$^{-1}$ about a factor $2$ 
smaller than the one by Dover et al.~\cite{Dover:1983}. 

In the present paper we report on a calculation of the deuteron annihilation
lifetime, considering neutron-antineutron oscillations. The main motivation comes, 
of course, from the aforementioned discrepancy and the work aims at providing if 
possible a plausible explanation for the difference. 
In our study we follow closely the benchmark calculations of
Sandars~\cite{Sandars:1980} and Dover et al.~\cite{Dover:1983}. 
The essential new aspect is that we employ modern interactions for the involved
$NN$ and the $\nnbar$ systems. Modern means that these interactions have
been derived in a consistent and systematic framework, namely chiral effective field 
theory (EFT) \cite{Epelbaum:2009}. And it means that these potentials are in line with 
present-day empirical information. This concerns specifically the $\nnbar$ system 
where a wealth of data on $\ppbar$ scattering and the charge-exchange reaction 
$\ppbar \to \nb n$ has accumulated in the years after the publication of Ref.~\cite{Dover:1983}, 
notably due to measurements at the LEAR facility at CERN \cite{Klempt:2002}.  
That fact is accounted for by utilizing $\nnbar$ potentials which have been fitted to 
up-to-date phase shifts and inelasticities provided by a recently 
published phase-shift analysis of available $\ppbar$ scattering data \cite{Zhou:2012}.

The $\nnbar$ interactions used in the present study are taken from two works~\cite{Kang:2013,Dai:2017}. 
In the first reference, $\nnbar$ potentials up to next-to-next-to-leading order (N$^2$LO) 
were constructed, based on a modified Weinberg power counting, in close analogy to 
pertinent studies of the $NN$ interaction \cite{Epelbaum:2005}.
In the second, the study was extended to next-to-next-to-next-to-leading order 
(N$^3$LO) and, in addition, a new regularization scheme was implemented that 
had been introduced in the $NN$ study of Ref.~\cite{Epelbaum:2015}.
In the actual calculations we rely mostly on the recent more refined potential 
of higher order \cite{Dai:2017} that describes $\nnbar$ phase shifts and scattering
observables up to laboratory energies of $T_{lab}\sim 250$~MeV. However, additional 
calculations with the other potential are performed in order to estimate possible 
uncertainties that arise, e.g., from the employed regularization scheme.  

The paper is structured in the following way: In Sect.~\ref{Formalism} a
basic description of the employed formalism is provided. Our results
are presented and discussed in Sect.~\ref{Results}. We compare also with 
the works of \cite{Dover:1983} and \cite{Oosterhof:2019}. 
The paper closes with a brief summary. 

\section{Formalism}
\label{Formalism}
In our treatment of $\osc$ oscillations in the deuteron we follow very closely the 
formalism outlined in Refs.~\cite{Sandars:1980,Dover:1983}.
However, contrary to those works our calculation is performed in momentum 
space and, therefore, we provide details about the main steps below. The 
starting point is the eigenvalue (Schr\"odinger) equation \cite{Sandars:1980} 
\begin{equation} 
 \begin{pmatrix}
  H_0 + V_{np}  & V_{\osc}\\
  V_{\osc} & H_0 + V_{\npbar} \\
 \end{pmatrix}
 \begin{pmatrix}
  |\psi_{np} \rangle \\
  |\psi_{\npbar} \rangle \\
 \end{pmatrix}
= (E - i \Gamma /2 ) 
 \begin{pmatrix}
  |\psi_{np} \rangle \\
  |\psi_{\npbar} \rangle \\
 \end{pmatrix} \ . 
\label{eq:EW}
\end{equation}
Here, $V_{np}$ and $V_{\npbar}$ are the potentials in the $np$ and $\bar n p$
systems and $|\psi_{np} \rangle$ and $|\psi_{\npbar} \rangle$ are the corresponding
wave functions. The systems are coupled via $V_{\osc}$ which is given by 
the ${\osc}$ transition matrix element $\delta m_{\osc}$ where the latter 
is proportional to the inverse of the $\osc$ oscillation time,
i.e. $V_{\osc} = \delta m_{\osc} = \hbar / \tau_{\osc}$ \cite{Phillips:2014}. 

To leading order the decay width of the deuteron, $\Gamma_d$, is \cite{Sandars:1980} 
\begin{equation} 
\Gamma_d = -2\, V_{\osc}\, {\rm Im} \langle \psi_{d} |\psi_{\npbar} \rangle \ , 
\label{eq:Gamma}
\end{equation} 
where $|\psi_{d} \rangle$ is the deuteron wave function. The wave 
function $|\psi_{\npbar} \rangle$ obeys the equation 
\begin{equation} 
(H_0 + V_{\nb p} - E) |\psi_{\npbar} \rangle = -V_{\osc} |\psi_{d} \rangle \ . 
\label{eq:LS}
\end{equation} 
We solve Eq.~(\ref{eq:LS}) in momentum space. Performing a partial wave
decomposition and taking into account the coupling of the $^3S_1$ and
$^3D_1$ channels, the above integral equation reads
\begin{eqnarray} 
(2 E_p - 2 E_\kappa)\,\psi^{L}_{\npbar} (p) 
&&+ \sum_{L'}\int \frac{dq\,q^2}{(2\pi)^3} 
V^{L,L'}_{\nb p}(p,q)\, \psi^{L'}_{\npbar} (q) \nonumber \\
&& = -V_{n-\nb}\,  \psi^L_{d}(p) \ , 
\label{eq:LSP}
\end{eqnarray} 
with $L,\,L' = 0,\,2$. 
Here, $2 E_\kappa$ is the total energy corresponding to the deuteron, i.e.
$2 E_\kappa - 2 m_N = 2 \sqrt{ m_N^2 - \kappa^2} - 2 m_N =-B_d$ 
where $B_d$ is the standard binding energy of $2.225$ MeV and $\kappa = \sqrt{m_N B_d} \simeq 45.7$~MeV
is the binding momentum. The deuteron wave function is normalized by
\begin{equation} 
\int dp p^2\, \left[(\psi^0_{d}(p))^2 + \psi^2_{d}(p))^2\right] = 1 \ , 
\label{eq:DWF}
\end{equation} 
and the width is provided by 
\begin{equation} 
\Gamma_d = -2\, V_{n-\nb}\, {\rm Im} \sum_L \int dp p^2\, \psi^L_{d}(p) \, \psi^L_{\npbar}(p) \ .
\label{eq:Width}
\end{equation} 
The deuteron lifetime $\tau_d$ is given by $\tau_d=\hbar /\Gamma_d$.
The interesting quantity is the so-called reduced lifetime $R$ 
\cite{Sandars:1980,Dover:1983,Friedman:2008} which relates the free $\osc$ 
oscillation lifetime with that of the deuteron, 
\begin{equation} 
\tau_d =  R\, \tau^2_{\osc} \quad {\rm i.e.} \quad
R = \frac{\hbar} {\Gamma_d \tau^2_{\osc} } \ . 
\label{eq:TR}
\end{equation}

\section{Results and discussion}
\label{Results}

As already stated in the Introduction, we use the $\bar NN$ interactions 
from Refs.~\cite{Kang:2013,Dai:2017} derived
within chiral EFT and the deuteron wave functions from the corresponding 
$NN$ potentials \cite{Epelbaum:2005,Epelbaum:2015}, derived in the same
framework. 
We consider the two $\nnbar$ interaction because they are based on
rather different regularization schemes. In the earlier potential
\cite{Kang:2013} a non-local exponential exponential regulator 
was employed for the whole potential while in the recent work \cite{Dai:2017} 
a local regulator was adopted for the evaluation of the one- and two-pion 
contributions, see Refs.~\cite{Dai:2017,Epelbaum:2015} for details. 
Comparing the pertinent results will allow us to shed light on the question
in how far the choice of the regulator influences the predictions. 
For exploring further the sensitivity of the results to the deuteron wave 
function we employ also those of two meson-exchange potentials 
\cite{Haidenbauer:1993,Machleidt:2001}. 
 
Our results are summarized in Table~\ref{tab:T1}. They are based on
our N$^3$LO interaction with cutoff $R_0 = 0.9$~fm from Ref.~\cite{Dai:2017}
and the N$^2$LO interaction with cutoff $\{\Lambda,\tilde\Lambda\} = 
\{450,500\}$~MeV from Ref.~\cite{Kang:2013}. For details on those interactions
we refer the reader to the corresponding publications. 
Besides the predictions for $R$ based on the chiral $\bar NN$ interactions 
we list also the values given in Ref.~\cite{Dover:1983} where the $\bar NN$ 
potentials DR$_1$ and DR$_2$ by Dover-Richard~\cite{Dover:1980,Dover:1982}
have been utilized. Furthermore we include results from the calculation of 
Oosterhof et al. performed directly within EFT on the basis of the KSW 
approach. In this case $R$ can be represented in a compact analytical 
form which reads up to NLO~\cite{Oosterhof:2019}
\begin{equation}  
R = -\frac{\kappa}{m_N}\, \frac{1}{{\rm Im}\,a_{\npbar}}\, 
\frac{1} {1+ 0.4+2 \kappa\, {\rm Re}\,a_{\npbar}-0.13\pm 0.4} \ . 
\label{eq:Oost}
\end{equation}
Obviously, the only parameter here is the $\nb p$ $^3S_1$ scattering length. 
All other quantities that enter are well established $NN$ observables, cf. 
Ref.~\cite{Oosterhof:2019} for details.  Note that in that paper, the scattering length
${\rm Re}\,a_{\npbar}$ was taken from Ref.~\cite{Dai:2017}.

\begin{table*}[ht]
\caption{Reduced lifetime ${R}$ calculated for the $\chi$EFT $\nnbar$ 
potentials from Refs.~\cite{Kang:2013,Dai:2017}, together with information 
on the pertinent $\nb p$ $^3S_1$ scattering length. 
Results for the Dover-Richard potentials DR$_1$ and DR$_2$ 
are taken from Ref.~\cite{Dover:1983}. The corresponding scattering lengths 
are from Ref.~\cite{Carbonell:1992}. 
Predictions based on Eq.~(\ref{eq:Oost}), i.e. on the KSW approach
applied in Ref.~\cite{Oosterhof:2019}, are indicated too. 
}
\label{tab:T1}
\vskip 0.1cm
\renewcommand{\arraystretch}{1.4}
\begin{center}
\begin{tabular}{|c|cc|cc|}
\hline
\hline
& $\chi$EFT N$^2$LO \cite{Kang:2013}  & $\chi$EFT N$^3$LO \cite{Dai:2017} 
  & DR$_1$ \cite{Dover:1983} & DR$_2$ \cite{Dover:1983} \\
\hline
\hline
${R}$ [s$^{-1}$] & $2.49\times 10^{22}$ & $2.56\times 10^{22}$ 
  & $2.56\times 10^{22}$ & $2.40\times 10^{22}$\\
(Eq.~(\ref{eq:Oost})) & $(1.1\pm 0.3)\times 10^{22}$ & $(1.2\pm 0.3)\times 10^{22}$ 
  & $(1.4\pm 0.4)\times 10^{22}$ & $(1.3\pm 0.3)\times 10^{22}$ \\
\hline
\hline
${a_{^3S_1}}$ [fm] & $0.44-{\rm i}\,0.91$ & $0.44-{\rm i}\,0.96$ 
  & $0.87-{\rm i}\,0.66$ & $0.89-{\rm i}\,0.71$ \\
\hline
\hline
\end{tabular}
\end{center}
\renewcommand{\arraystretch}{1.0}
\end{table*}

As can be seen from Table~\ref{tab:T1} the values for $R$ predicted by 
the chiral $\bar NN$ interactions are fairly similar to those obtained
for the DR potentials in the past. The values based on the framework 
utilized by Oosterhof et al.~\cite{Oosterhof:2019}, on the other 
hand, are rather different as already pointed out by the authors of that work. 
The large discrepancy is definitely not due to differences in Im~$a_{\nb p}$
as suggested in Ref.~\cite{Oosterhof:2019} where the scattering length
from our N$^3$LO chiral $\bar NN$ interaction \cite{Dai:2017} was used, 
but clearly a consequence of the different approaches employed. 

\begin{figure}[h]
\begin{center}
 \includegraphics[height=65mm,angle=0]{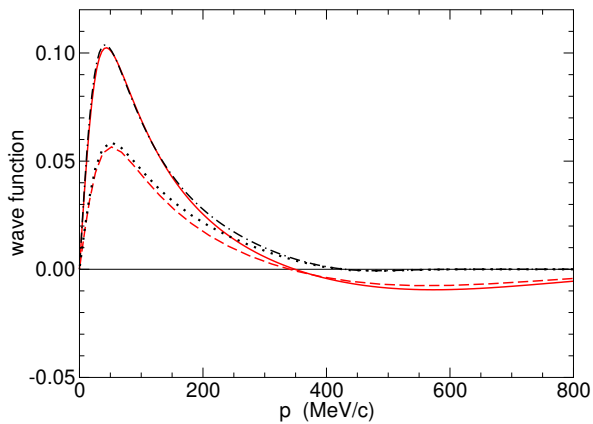}
\caption{Wave functions of the deuteron ($\psi_d$; upper curves) and of 
the $\bar n p$ state in the $^3S_1$ partial wave (imaginary part 
of $\psi_{\bar n p}$; lower curves), multiplied by the momentum $p$ 
and with arbitrary normalization.
Solid and dashed (red) curves are for the N$^3$LO $\nnbar$ interaction \cite{Dai:2017}
with cutoff $R_0=0.9$~fm and 
dash-dotted and dotted (black) curves for the N$^2$LO $\nnbar$ interacton \cite{Kang:2013}
with cutoff $\{\Lambda,\tilde\Lambda\}=\{450,500\}$~MeV. 
}
\label{fig:3S1}
\end{center}
\end{figure}

For illustration we present in Fig.~\ref{fig:3S1} the wave functions
of the $^3S_1$ component of the deuteron and of the corresponding
$\bar n p$ state with arbitrary normalizations. In case of the latter
the imaginary part is shown which is relevant for the determination of 
the width, cf. Eq.~(\ref{eq:Width}). The impact of the different 
regularization schemes used in Refs.~\cite{Epelbaum:2005,Kang:2013} and 
\cite{Epelbaum:2015,Dai:2017}, respectively, can be clearly seen from the 
different behavior of the wave functions for large momenta. Note, however, 
that the bulk contribution to the integral in Eq.~(\ref{eq:Width}) comes 
from momenta $p\le 300$~MeV/c. Contributions from larger momenta 
to $\Gamma_d$ are only in the order of $5$\,\% as revealed by test
calculations. 

In order to investigate the sensitivity of our results to the utilized
input we performed various exploratory calculations. First of all, we 
employed the NLO and N$^2$LO variants of the considered $\nnbar$ (and $NN$)
interactions. The corresponding predictions for $R$ were found to lie 
within a range of $(2.48 - 2.65)\times 10^{22}$~s$^{-1}$. 
Taking this variation as measure for the uncertainty due to the nuclear structure, 
i.e. the $NN$ and $\nnbar$ interactions (wave functions), leads to a value of 
roughly $R = (2.6\pm 0.1)\times 10^{22}$~s$^{-1}$. Applying the method 
proposed in Ref.~\cite{Epelbaum:2015} for estimating the uncertainty to the 
calculation based on the $\nnbar$ interaction from 2017 \cite{Dai:2017}, 
say, would actually lead to a slightly smaller uncertainty. 
We have also varied the deuteron wave functions alone. As an extreme case 
we even took wave functions from phenomenological $NN$ potentials derived 
in an entirely different framework, namely in the meson-exchange 
picture \cite{Haidenbauer:1993,Machleidt:2001}. Also here the obtained
values for $R$ stayed within the range given above. 
Finally, omitting the $D$-wave component of the deuteron wave function, which 
is kept in our calculation and in the one by Dover et al.~\cite{Dover:1983} 
causes a $5$~\% variation. But it leads to an increase of the value of $R$ 
and, thus, does not bring it closer to the values presented by Oosterhof 
et al.  
Overall, we confirm the observation by Dover et al. that the predictions for
$R$ are fairly insensitive to the details of the employed $\nnbar$
potentials \cite{Dover:1983}, provided that these potentials describe 
the $\bar pp$ data at low energies. 

Since in Ref.~\cite{Oosterhof:2019} the $\nnbar$ interaction is represented
by the leading-order term, i.e. the scattering length alone, we have 
evaluated the effective range parameter for our $\nnbar$ interactions and 
used it to extrapolate the $\nnbar$ amplitude to the deuteron pole.  
The found variations are in the order of $10$~\% and can 
certainly not explain the large difference documented in Table~\ref{tab:T1}. 

We do not have a clear explanation for the difference of our results 
(and those of Ref.~\cite{Dover:1983}) to the ones of Oosterhof et 
al.~\cite{Oosterhof:2019}. However, we believe that it is due to the
fact that in the latter work the width $\Gamma_d$ is evaluated following the 
perturbative scheme developed by Kapan, Savage, and Wise \cite{Kaplan:1999}. 
In that scheme there is no proper deuteron wave function. Rather 
one works with an effectively constructed wave function that is represented 
in terms of an irreducible two-point function~\cite{Oosterhof:2019,Kaplan:1999}. 
This seems to work well for some electromagnetic form factors of
the deuteron, at least at low momentum transfer~\cite{Kaplan:1999,Walzl:2001}. 
On the other hand, the quadrupole moment of the deuteron is 
overestimated by $40$\,\% \cite{Kaplan:1999}, which hints that the 
properties of the wave function at large distances (small momenta) are 
not that well represented in this scheme. Clearly, this should have an 
impact on the quantity studied in the present work as well. 
Note that a comparable agreement (mismatch) with regard to the KSW scheme 
has been also observed in studies of the electric dipole moment (magnetic 
quadrupole moment) of the deuteron~\cite{deVries:2011,Liu:2012,Bsaisou:2015}. 
In any case, one should not forget that there is convergence problem of the 
KSW approach for $NN$ partial waves where the tensor force from pion exchange 
is present~\cite{Fleming:2000}. It affects specifically the $^3S_1$-$^3D_1$ 
channel where difficulties appear already for momenta around $100$~MeV/c, 
see~\cite{Fleming:2000} and also the discussions in 
Refs.~\cite{Epelbaum:2009,Bedaque:2002}. 
Interestingly enough, this is the momentum region where the dominant 
contributions in our calculation come from, see Fig.~\ref{fig:3S1}. 
 
In contrast to \cite{Oosterhof:2019} in our study the $\nnbar$ ($NN$) 
and $|\Delta B| =2$ interactions are not treated at the same order. 
Guided by the success of the work by Epelbaum et al.
\cite{Epelbaum:2005,Epelbaum:2015} based on the Weinberg counting
the non-perturbative effects due to the $NN$ and $\nnbar$ 
interactions are fully take into account from the very beginning. 
It is worth mentioning that in this approach the quadrupole moment 
of the deuteron is very close to its empirical value already for 
interactions at NLO \cite{Epelbaum:2005,Epelbaum:2015}. 

Finally, note that in the present work we have only considered the contribution 
from actual $\osc$ oscillations to the deuteron decay rate. In principle, there 
can be also contributions from a direct annihilation of the $NN$ system. 
Moreover, there can be a $NN\to \nnbar$ transition involving both nucleons  
and the $\nnbar$ state. The latter gives rise to a 
$|\Delta B| = 2$ four-baryon contact term involving an unknown complex 
low-energy constant, which has been discussed and considered in the work by 
Oosterhof et al.~\cite{Oosterhof:2019}. We refrain from introducing such 
a contribution in the present study which, anyway, is of higher 
order in the employed counting scheme. However, we would like to mention 
that according to Ref.~\cite{Oosterhof:2019} the deuteron decay rate  
is indeed dominated by free $\osc$ oscillations. Nonetheless, an 
uncertainty in the order of $\pm 0.3 \times 10^{22}$~s$^{-1}$ for the 
reduced lifetime $R$ due to additional $|\Delta B|=2$ mechanisms, as 
suggested in that work, has to be certainly expected. 
Note that additional decay mechanisms should always increase the decay width of 
the deuteron and, thus, lead to smaller values of the reduced lifetime $R$. 

\section{Summary}
In the present work we have considered neutron-antineutron 
oscillations in the deuteron. In particular, we have calculated 
the deuteron lifetime in terms of the free-space $\osc$ oscillation time
based on $\nnbar$ \cite{Kang:2013,Dai:2017} and 
$NN$ \cite{Epelbaum:2005,Epelbaum:2015} interactions derived within 
chiral effective field theory. The value obtained for the 
so-called reduced lifetime $R$ which relates the free-space $\osc$
oscillation time $\tau_{\osc}$ with the deuteron lifetime is found to
be $R=(2.6\pm 0.1) \times 10^{22}$~s$^{-1}$, where the quoted uncertainty is due
to the $NN$ and $\nnbar$ interactions (wave functions). 

Our result for $R$ is close to the value obtained by Dover and collaborators
more than three decades ago \cite{Dover:1983} but disagrees with recent EFT 
calculations, performed within the perturbative scheme proposed by Kaplan, Savage, 
and Wise \cite{Oosterhof:2019}, by about a factor of~$2$.
We believe that the difference is due to the fact that in the perturbative 
KSW scheme there is no proper deuteron wave function. Rather this ingredient
is represented effectively in terms of an irreducible two-point function. 
It is known from past studies that the KSW approach fails to describe quantities 
that depend more sensitively on the wave function like, for example, the quadrupole 
moment of the deuteron \cite{Kaplan:1999}.  

\section*{Acknowledgements}
We acknowledge stimulating discussions with Jordy de Vries, Andreas Nogga and Tom Luu.
This work is supported in part by the DFG and the NSFC through
funds provided to the Sino-German CRC 110 ``Symmetries and
the Emergence of Structure in QCD'' (DFG grant. no. TRR~110)
and the VolkswagenStiftung (grant no. 93562).
The work of UGM was supported in part by The Chinese Academy
of Sciences (CAS) President's International Fellowship Initiative (PIFI) (grant no.~2018DM0034).

\end{document}